\documentclass[conference]{IEEEtran}
\IEEEoverridecommandlockouts
\usepackage{booktabs}
\usepackage{adjustbox}
\usepackage{cite}
\usepackage{multirow}
\usepackage{amsmath,amssymb,amsfonts}
\usepackage{tabularx}
\usepackage{bm}
\usepackage{algorithmic}
\usepackage{graphicx}
\usepackage{textcomp}
\usepackage{xcolor}
\def\BibTeX{{\rm B\kern-.05em{\sc i\kern-.025em b}\kern-.08em
    T\kern-.1667em\lower.7ex\hbox{E}\kern-.125emX}}

\author{\IEEEauthorblockN{Anonymous Authors}
}

\begin{document}

\title{S$2$SCORE: A Textless Unified Evaluation Metric for Speech-to-Speech Translation Systems}
\author{
\IEEEauthorblockN{
Lalaram Arya$^{\star}$ \qquad
Mrinmoy Bhattacharjee$^{\dagger}$ \qquad
S. R. Mahadeva Prasanna$^{\star}$
}
\IEEEauthorblockA{
$^{\star}$\textit{Department of EECE, IIT Dharwad, Dharwad, India} \\
\texttt{202021004@iitdh.ac.in}, \texttt{prasanna@iitdh.ac.in}
\\[2pt]
$^{\dagger}$\textit{Department of CSE, IIT Jammu, Jammu, India} \\
\texttt{mrinmoy.bhattacharjee@iitjammu.ac.in}
}
}

\maketitle

\begin{abstract}
Evaluating speech-to-speech translation (S$2$ST) systems requires assessing multiple dimensions of translated speech, including semantic adequacy, speaker similarity, and acoustic naturalness. Existing approaches are largely text-based, relying on ASR transcripts and primarily assessing semantic adequacy, while other dimensions require separate metrics. We propose S$2$SCORE, a unified text-free evaluation framework that jointly models these dimensions within a multi-task architecture and produces a human-calibrated overall quality score. S$2$SCORE is trained using supervision derived from established evaluation models, enabling scalable learning without direct human annotations. Evaluated across five language pairs, S$2$SCORE shows strong agreement with human judgments, independent reference metrics, and the training supervision metrics, indicating robust generalization. The reference-free variant, S$2$SCORE-QE, consistently outperforms S$2$SCORE-REF and achieves stronger semantic adequacy prediction than BLASER~$2.0$.

\end{abstract}


\section{Introduction}
Speech-to-Speech Translation (S$2$ST)~\cite{dhawan2022s2st} enables spoken communication across languages by translating source-language speech into target-language speech. Recent advances in S$2$ST~\cite{Etchego2022} have increasingly shifted toward end-to-end architectures that avoid intermediate Automatic Speech Recognition (ASR)~\cite{alharbi2021asr}, Machine Translation (MT)~\cite{tan2020nmt}, and Text-to-Speech (TTS)~\cite{casanova2024xtts} components, reducing latency and error propagation. However, removing intermediate text representations also introduces new challenges for evaluation.

Translation quality has traditionally been assessed using text-based metrics such as BLEU~\cite{papineni2002bleu}, METEOR~\cite{banerjee2005meteor}, BLEURT~\cite{sellam2020bleurt}, COMET~\cite{rei2020comet}, and MetricX-$24$~\cite{juraska-etal-2024-metricx}. Modern ASR systems~\cite{radford2023whisper} have made ASR-based evaluation a strong baseline for semantic evaluation. However, because these approaches rely solely on text, they discard acoustic information relevant to translation quality assessment. They also require an additional ASR system, introducing computational overhead and potential transcription errors.

BLASER~\cite{blaser} introduced one of the first speech-based metrics supervised with human judgments for semantic adequacy, while BLASER~2.0~\cite{blaser2} extended it to massively multilingual and reference-free evaluation. In parallel, a textless S$2$ST metric~\cite{besacier2023textless} was proposed to approximate text-based metrics without relying on transcripts, while SpeechQE~\cite{speechqe} recently proposed a Large Language Model (LLM)-based framework for Speech Translation (ST) quality estimation.

Despite these advances, existing text-free S$2$ST evaluation methods remain largely focused on semantic adequacy. In modern S$2$ST systems~\cite{le2024transvip}, output quality is multidimensional: translations should not only preserve meaning but also maintain speaker identity and naturalness. Consequently, evaluation relies on separate metrics for each dimension, leading to fragmented pipelines and increased evaluation overhead~\cite{jia2019direct}. To the best of our knowledge, no existing text-free evaluation framework jointly models these dimensions and provides a unified overall quality score.

To address these limitations, we propose S$2$SCORE, a unified text-free evaluation framework that jointly models semantic adequacy, speaker similarity, and acoustic naturalness within a multi-task architecture while producing a learned overall quality score calibrated against human judgments. Since collecting large-scale human judgments is costly and labor-intensive, the framework is trained using supervision derived from established evaluation metrics: xCOMET-XXL~\cite{xcomet} for semantic adequacy, WavLM-SV~\cite{chen2022wavlm} for speaker similarity, and UTMOSv$2$~\cite{baba2024utmosv2} for acoustic naturalness. S$2$SCORE leverages SONAR~\cite{duquenne2023sonar}, WavLM-SV~\cite{chen2022wavlm}, and log-Mel spectrogram~\cite{davis1980comparison} features for scalable multilingual evaluation. Training and evaluation use multilingual S$2$ST data derived from the CVSS~\cite{jia2022cvss} and BhasaAnuvaad~\cite{jain2025bhasaanuvaad} corpora, covering French (fr), Spanish (es), German (de), Chinese (zh), and Hindi (hi) speech translated into English (en) and generated by both cascaded and direct S$2$ST systems, yielding diverse translation quality conditions. Experimental results show that S$2$SCORE provides an effective multidimensional speech quality evaluation, achieving strong agreement with human judgments and independent reference metrics. The key contributions are:

\begin{figure*}[!t]
\centering
\includegraphics[width=0.80\textwidth]{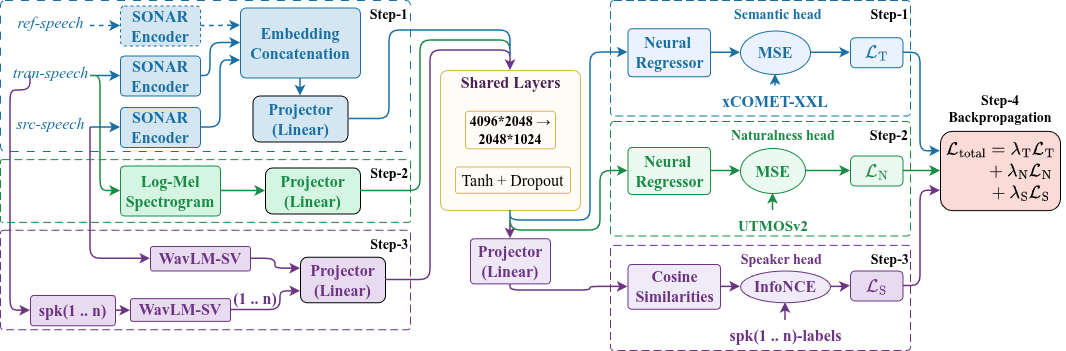}
\caption{Architecture of S$2$SCORE. Semantic, acoustic, and speaker representations are projected into a shared MLP and processed by task-specific heads supervised by xCOMET-XXL~\cite{xcomet}, UTMOSv$2$~\cite{baba2024utmosv2}, and WavLM-SV~\cite{chen2022wavlm}. Dashed arrows denote reference speech used only in the reference-based variant.}
\label{s3}
\end{figure*}
\begin{itemize}
\item We introduce S$2$SCORE, a unified text-free evaluation framework for S$2$ST that jointly models semantic adequacy, speaker similarity, and acoustic naturalness and produces a human-calibrated overall quality score.

\item We propose a scalable supervision strategy that leverages established metrics to train multidimensional speech evaluation models without large-scale human annotation.

\item We evaluate S$2$SCORE across five language pairs, demonstrating strong agreement with human judgments and independent reference metrics.
\end{itemize}



\section{Proposed S2SCORE Framework}
\label{sec:proposed_sec}
S$2$SCORE adopts a unified multi-head architecture to evaluate semantic adequacy, speaker similarity, and acoustic naturalness in S$2$ST systems while producing a unified overall quality score. As illustrated in Fig.~\ref{s3}, the framework comprises task-specific feature extractors, a shared MLP module (S-MLP), and three prediction heads. Input speech is encoded into task-specific representations, projected into a common embedding space, and processed by the shared S-MLP before being routed to the corresponding heads. The resulting semantic adequacy, speaker similarity, and naturalness scores are subsequently aggregated into the unified overall quality score. The individual components are described below.

\subsection{Speech Feature Representations}
\label{subsec:represent}
S$2$SCORE evaluates translated speech along three dimensions, each represented using a task-specific feature extractor. Semantic representations are extracted using the SONAR encoder~\cite{duquenne2023sonar}, which provides multilingual speech and text embeddings as it is trained on $57$ speech and over $200$ text languages, making it well-suited for cross-lingual evaluation. Acoustic naturalness is represented using log-Mel spectrograms~\cite{davis1980comparison}, which capture perceptually relevant spectro-temporal characteristics. Speaker similarity is modeled using pretrained WavLM-SV embeddings~\cite{chen2022wavlm}, a speaker verification model trained on approximately $94$k hours of speech that provides robust speaker representations.

\subsection{Shared MLP Layers}
\label{subsec:sharedlayer}
Following the hard parameter-sharing approach~\cite{Yu2025Multitask}, the semantic, speaker, and naturalness heads share a common S-MLP module. The S-MLP maps a $4096$-dimensional input to a shared representation $\mathbf{h}\in\mathbb{R}^{1024}$ through two fully connected layers with $2048$ and $1024$ units, each followed by \emph{tanh} activation and dropout ($p=0.1$). The resulting representation is passed to the corresponding prediction heads, and the S-MLP parameters are jointly optimized using the weighted sum of the three task losses.

\subsection{Evaluation Heads}
\label{evalhead}
The evaluation heads operate on the output of the S-MLP (Section~\ref{subsec:sharedlayer}) to predict task-specific scores.

\subsubsection{Translation Evaluation Head}
The translation head estimates semantic adequacy using SONAR embeddings~\cite{duquenne2023sonar}. Two variants of S$2$SCORE are developed, S$2$SCORE-QE and  S$2$SCORE-REF, depending on reference speech availability. Following BLASER~2.0~\cite{blaser2}, interaction features are constructed using element-wise products and absolute differences between source and translated embeddings, with reference embeddings additionally included in the REF variant. The resulting features are projected and processed by the S-MLP to produce $\mathbf{h}_{\mathrm{T}} \in \mathbb{R}^{1024}$. Subsequently, the translation quality score is predicted by a regression head $f_{\mathrm{T}}(\cdot)$ as follows:
\begin{equation}
    \hat{S2SCORE}_{\mathrm{T}} = f_{\mathrm{T}}(\mathbf{h}_{\mathrm{T}})
    \label{eq:tran_pred}
\end{equation}
The xCOMET-XXL~\cite{xcomet} scores $\hat{y}_{\mathrm{xCOMET-XXL}}$ are used as supervision for computing Mean Squared Error (MSE) loss as follows:

\begin{equation}
    \mathcal{L}_{\mathrm{T}} =
    \lVert\hat{S2SCORE}_{\mathrm{T}} - \hat{y}_{\mathrm{xCOMET-XXL}}\rVert^2
    \label{eq:loss_tran}
\end{equation}

\subsubsection{Acoustic Naturalness Head}
\label{naturalness}
The naturalness head estimates the perceptual quality of translated speech from log-Mel spectrogram features~\cite{davis1980comparison}, following UTMOSv$2$~\cite{baba2024utmosv2}. Log-Mel spectrograms $\mathbf{M} \in \mathbb{R}^{T \times F}$ are extracted from the translated signal, flattened, and mapped by a task-specific projector into the common $4096$-dimensional space. The projected representation is then processed by the S-MLP to produce $\mathbf{h}_{\mathrm{N}} \in \mathbb{R}^{1024}$, which is passed to a regression head $f_{\mathrm{N}}(\cdot)$:
\begin{equation}
    \hat{S2SCORE}_{\mathrm{N}} = f_{\mathrm{N}}(\mathbf{h}_{\mathrm{N}})
\end{equation}
The head is optimized using MSE against UTMOSv$2$~\cite{baba2024utmosv2} scores $\hat{y}_{\mathrm{UTMOSv2}}$:
\begin{equation}
\mathcal{L}_{\mathrm{N}} =
\lVert \hat{S2SCORE}_{\mathrm{N}} - \hat{y}_{\mathrm{UTMOSv2}}\rVert^2
\label{eq:loss_nat}
\end{equation}

\subsubsection{Speaker Similarity Head}
\label{speaker-sim}
To evaluate speaker preservation, the source utterance is paired with $N$ translated candidates, comprising one positive match and $N-1$ negatives. Speaker embeddings $\mathbf{e}^{\mathrm{sv}}_{\mathrm{src}}$ and $\mathbf{e}^{\mathrm{sv}}_{\mathrm{tran},i}$ are extracted using WavLM-SV~\cite{chen2022wavlm}, projected into the shared $4096$-dimensional feature space, and processed by the S-MLP to obtain $\mathbf{h}_{\mathrm{S}} \in \mathbb{R}^{1024}$. A final projection maps these representations to $128$-dimensional embeddings, $\tilde{\mathbf{e}}^{\mathrm{sv}}_{\mathrm{src}}$ and $\tilde{\mathbf{e}}^{\mathrm{sv}}_{\mathrm{tran},i}$, and the contrastive score is computed using cosine similarity, as defined below:

\begin{equation}
\hat{S2SCORE}_{\mathrm{S}}^{i} =
\frac{
\tilde{\mathbf{e}}^{\mathrm{sv}}_{\mathrm{src}} \cdot
\tilde{\mathbf{e}}^{\mathrm{sv}}_{\mathrm{tran},i}
}{
\lVert \tilde{\mathbf{e}}^{\mathrm{sv}}_{\mathrm{src}} \rVert
\lVert \tilde{\mathbf{e}}^{\mathrm{sv}}_{\mathrm{tran},i} \rVert
}
\end{equation}
The head is trained using InfoNCE loss~\cite{oord2018representation} as follows:
\begin{equation}
\mathcal{L}_{\mathrm{S}} =
-\log
\frac{\exp(\hat{S2SCORE}_{\mathrm{S}}^{+}/\tau)}
{\sum_{i=1}^{N} \exp(\hat{S2SCORE}_{\mathrm{S}}^{i}/\tau)}
\end{equation}
where $\hat{S2SCORE}_{\mathrm{S}}^{+}$ is the similarity score of the matched source-translation pair, $\hat{S2SCORE}_{\mathrm{S}}^{i}$ denotes the score of the $i^{th}$ candidate, and $\tau$ is the temperature parameter.

\subsubsection{Unified Training Objective}

The overall training objective is the weighted sum of the three task-specific losses:
\begin{equation}
\mathcal{L}_{\mathrm{total}} =
\lambda_{\mathrm{T}} \mathcal{L}_{\mathrm{T}} +
\lambda_{\mathrm{N}} \mathcal{L}_{\mathrm{N}} +
\lambda_{\mathrm{S}} \mathcal{L}_{\mathrm{S}}
\end{equation}
The weights $\lambda_{\mathrm{T}}$, $\lambda_{\mathrm{N}}$, and $\lambda_{\mathrm{S}}$ are selected on the validation set to balance the three losses during joint optimization.

\subsubsection{Unified Speech Quality Score}
\label{subsec:unified_score}
Beyond the individual scores, S$2$SCORE produces a unified speech quality score that integrates semantic adequacy, acoustic naturalness, and speaker similarity into a single interpretable value. Since the three heads operate on different ranges ($\hat{S2SCORE}_{\mathrm{T}} \in [0,1]$, $\hat{S2SCORE}_{\mathrm{N}} \in [1,5]$, and $\hat{S2SCORE}_{\mathrm{S}} \in [-1,1]$), the scores are normalized to $[0,1]$. The unified score is computed as a weighted combination of the normalized scores ($\tilde{\cdot}$):

\begin{equation}
\begin{split}
S2SCORE
=&\;
\alpha_{\mathrm{T}}\,\tilde{S2SCORE}_{\mathrm{T}}
+
\alpha_{\mathrm{N}}\,\tilde{S2SCORE}_{\mathrm{N}}
\\
&+
\alpha_{\mathrm{S}}\,\tilde{S2SCORE}_{\mathrm{S}}
\end{split}
\label{eq:overall}
\end{equation}
where $\alpha_{\mathrm{T}} + \alpha_{\mathrm{N}} + \alpha_{\mathrm{S}} = 1$ and $S2SCORE \in [0,1]$. The aggregation weights are learned from human ratings of overall quality on a held-out validation set, yielding a unified score calibrated to human judgments of S$2$ST quality.

\section{Experimental Setup}
\label{sec:experiments}
This section discusses the datasets, training configuration, and evaluation metrics used to train and assess S$2$SCORE.

\subsection{Dataset}
\label{dataset}
Training S$2$SCORE requires source, translated, and reference speech with corresponding transcriptions for generating supervision labels (reference speech is required only for the REF variant). Existing S$2$ST corpora provide source and reference speech but lack sufficiently diverse translated speech spanning a broad range of semantic adequacy, speaker similarity, and acoustic naturalness.

To address this limitation, we derive a multilingual dataset from CVSS~\cite{jia2022cvss} and BhasaAnuvaad~\cite{jain2025bhasaanuvaad}. We use the \textit{fr--en}, \textit{de--en}, \textit{es--en}, and \textit{zh--en} language pairs from CVSS and the \textit{hi--en} pair from BhasaAnuvaad. CVSS provides source speech, transcripts, English translations, and synthesized reference speech (CVSS-T), while BhasaAnuvaad provides source and reference speech with aligned transcriptions. Translated speech is generated using multiple S$2$ST systems (Section~\ref{subsubsec:s2st_models}), introducing variation in translation quality. The resulting dataset is divided into training, validation, and test sets using an $80\%/10\%/10\%$ split, with no sentence or speaker overlap. Dataset statistics are summarized in Table~\ref{tab:evaldatatable}.

\begin{table}[!t]
\centering
\caption{Dataset statistics summarizing the number of sentences in thousands (k) and speech duration in hours (h) for the train, validation, and test splits across language pairs.}
\label{tab:evaldatatable}
\setlength{\tabcolsep}{2pt}
\renewcommand{\arraystretch}{1.15}
\begin{tabularx}{\columnwidth}{
  >{\centering\arraybackslash}p{1.4cm}
  >{\centering\arraybackslash}X
  >{\centering\arraybackslash}X
  >{\centering\arraybackslash}X
}
\hline
\textbf{Language} &
\textbf{Train} &
\textbf{Validation} &
\textbf{Test} \\
\cline{1-4}
fr--en &
\mbox{$11.8$k \textbar{} $18.4$h} &
\mbox{$1.5$k \textbar{} $2.3$h} &
\mbox{$1.5$k \textbar{} $2.2$h} \\
es--en &
\mbox{$10.6$k \textbar{} $9.8$h} &
\mbox{$1.3$k \textbar{} $1.2$h} &
\mbox{$1.3$k \textbar{} $1.2$h} \\
de--en &
\mbox{$10.8$k \textbar{} $9.7$h} &
\mbox{$1.4$k \textbar{} $1.2$h} &
\mbox{$1.4$k \textbar{} $1.1$h} \\
zh--en &
\mbox{$5.6$k \textbar{} $7.4$h} &
\mbox{$0.7$k \textbar{} $1.1$h} &
\mbox{$0.7$k \textbar{} $1.0$h} \\
hi--en &
\mbox{$9.6$k \textbar{} $11.8$h} &
\mbox{$1.2$k \textbar{} $1.6$h} &
\mbox{$1.2$k \textbar{} $1.5$h} \\
\hline
\end{tabularx}
\end{table}

\subsubsection{Translated Speech Generation}
\label{subsubsec:s2st_models}
To generate translated speech spanning a broad range of quality conditions, we employ three S$2$ST systems:

\begin{itemize}
\item A cascaded pipeline based on Whisper large-v$2$~\cite{radford2023whisper} for Speech Translation (ST), followed by speech synthesis using XTTS-v$2$~\cite{casanova2024xtts}.

\item SeamlessM4T-v$2$~\cite{barrault2023seamlessm4t}, used in both cascaded (ASR+MT+TTS) and ST modes, with XTTS-v$2$ for speech synthesis.

\item DS2ST-LM~\cite{arya2026timbreaware}, a direct S$2$ST model.
\end{itemize}
For each system, translated speech is synthesized using six target voices (three male and three female) per source utterance, enabling matched and mismatched speaker pairs for contrastive training of the speaker similarity head.

\begin{table}[!t]
\centering
\caption{Corpus-level ASR-BLEU~\cite{papineni2002bleu} and xCOMET-XXL~\cite{xcomet} scores used to assess dataset quality across language pairs.}
\label{tab:bleu_asr_comet}
\setlength{\tabcolsep}{5pt}
\renewcommand{\arraystretch}{1.15}
\begin{tabular}{lccccc}
\hline
\textbf{Metric} & \textbf{fr--en} & \textbf{es--en} & \textbf{de--en} & \textbf{zh--en} & \textbf{hi--en} \\
\hline
ASR-BLEU &
$26.57$ & $23.91$ & $20.12$ & $15.16$ & $23.48$ \\

xCOMET-XXL &
$0.587$ & $0.571$ & $0.554$ & $0.567$ & $0.591$ \\
\hline
\end{tabular}
\end{table}

\subsubsection{Data Quality}
\label{subsubsec:data_quality}
As shown in Table~\ref{tab:bleu_asr_comet}, ASR-BLEU varies substantially across language pairs ($15.16$--$26.57$), reflecting differences in translation difficulty, whereas xCOMET-XXL~\cite{xcomet} remains relatively stable ($0.554$--$0.591$), suggesting consistent semantic adequacy. The sentence-level xCOMET-XXL distribution in Fig.~\ref{fig:score_dist}(c) spans the full $[0,1]$ range without severe skew, indicating diverse translation quality.

\subsubsection{Supervision Label Generation}
\label{subsubsec:pseudo_labels}
Large-scale human annotation across multiple languages is costly and labor-intensive. Therefore, supervision labels for the individual evaluation dimensions are generated automatically at scale using established evaluation metrics. Semantic adequacy is supervised using xCOMET-XXL~\cite{xcomet} scores computed from Whisper large-v$2$~\cite{radford2023whisper} ASR transcripts of translated speech, acoustic naturalness using UTMOSv$2$~\cite{baba2024utmosv2}, and speaker preservation contrastively using the matched and mismatched speaker pairs described in Section~\ref{subsubsec:s2st_models}. 

Human ratings for aggregation-weight learning and evaluation were collected from five proficient non-native English speakers with graduate-level education in speech and language processing. To learn the aggregation weights $(\alpha_{\mathrm{T}}, \alpha_{\mathrm{N}}, \alpha_{\mathrm{S}})$, $2500$ translated utterances were randomly sampled from the training and validation sets ($500$ per language pair). Annotators rated overall quality on a five-point scale, considering semantic adequacy, acoustic naturalness, and speaker preservation, given the translated speech, a reference speech sample of the target speaker, and the English reference transcription. The ratings were averaged and linearly normalized to $[0,1]$ for weight learning. For evaluation, an additional $2500$ utterances ($500$ per language pair) were sampled from the held-out test set and independently rated for semantic adequacy, acoustic naturalness, speaker preservation, and overall quality using the same protocol. The five annotators achieved a Krippendorff's $\alpha$ of $0.79$, indicating substantial inter-annotator agreement.

\subsection{Training Setup}
\label{sec}

The SONAR~\cite{duquenne2023sonar} and WavLM-SV~\cite{chen2022wavlm} encoders remain frozen. For the naturalness branch, log-Mel spectrograms~\cite{davis1980comparison} with $F=64$ Mel-frequency bins are truncated or zero-padded to $T=256$ frames and flattened into a $16{,}384$-dimensional vector. The semantic and naturalness heads use two hidden layers with dimensions $[512, 64]$, ReLU activation, and dropout ($p=0.1$), while the speaker branch is trained using an InfoNCE~\cite{oord2018representation} objective with temperature $\tau=0.07$. The model is trained for $250$ epochs using AdamW~\cite{loshchilov2019adamw} with a learning rate of $5\times10^{-4}$, batch size $64$, and linear warm-up. Multi-task loss weights are set to $\lambda_{\mathrm{T}}=1.0$, $\lambda_{\mathrm{N}}=0.5$, and $\lambda_{\mathrm{S}}=0.1$, selected using the validation set. Aggregation weights $(\alpha_{\mathrm{T}}, \alpha_{\mathrm{N}}, \alpha_{\mathrm{S}})$ are subsequently learned from human overall quality ratings. All experiments are conducted on a single NVIDIA Quadro RTX~$6000$ ($24$ GB) GPU.


\subsection{Evaluation Metrics}
\label{subsec:eval_protocol}
We use Spearman's rank correlation~\cite{spearman1904proof} ($\rho$) as the primary evaluation metric. Correlations between S$2$SCORE predictions and human ratings are reported for semantic adequacy, acoustic naturalness, speaker preservation, and overall quality. To assess generalization and alignment, we report correlations with independent metrics (MetricX-$24$~\cite{juraska-etal-2024-metricx}, NISQA~\cite{mittag2021nisqa}, and ECAPA-TDNN~\cite{desplanques2020ecapa}) and training signals (xCOMET-XXL~\cite{xcomet}, UTMOSv$2$~\cite{baba2024utmosv2}, and WavLM-SV~\cite{chen2022wavlm}). Statistical significance is assessed using paired bootstrap resampling~\cite{koehn2004statistical}.

\section{Experimental Results and Analysis}
\label{sec:results}
This section presents the experimental evaluation of S$2$SCORE, results analysis, and ablation studies.

\begin{table}[!t]
\footnotesize
\centering
\caption{Spearman correlation ($\rho$) of semantic-quality predictions with human adequacy ratings, MetricX-$24$~\cite{juraska2023metricx}, and xCOMET-XXL~\cite{xcomet} scores on test split. Mean values are computed across the five language pairs. Best results within each reference category are shown in bold.}
\label{tab:semantic_eval}
\setlength{\tabcolsep}{5.0pt}
\renewcommand{\arraystretch}{1.05}

\begin{tabular}{lcccccc}
\toprule
\textbf{Model} &
\textbf{fr--en} &
\textbf{es--en} &
\textbf{de--en} &
\textbf{zh--en} &
\textbf{hi--en} &
\textbf{Mean} \\
\midrule

\multicolumn{7}{l}{\textit{Human adequacy ratings}} \\
\cmidrule(lr){1-7}
BLASER~$2.0$-REF
& 0.615 & 0.608 & 0.583 & 0.578 & 0.603 & 0.597 \\

BLASER~$2.0$-QE
& 0.619 & 0.613 & 0.597 & 0.591 & 0.610 & 0.606 \\

S$2$SCORE$_T$-REF
& 0.635 & 0.629 & 0.617 & 0.604 & 0.619 & 0.621 \\

S$2$SCORE$_T$-QE
& \textbf{0.649} & \textbf{0.637} &
  \textbf{0.628} & \textbf{0.607} &
  \textbf{0.633} & \textbf{0.631} \\
\midrule

\multicolumn{7}{l}{\textit{MetricX-$24$ (Independent semantic metric)}} \\
\cmidrule(lr){1-7}
S$2$SCORE$_T$-REF
& 0.648 & 0.637 & 0.625 & 0.610 & 0.634 & 0.631 \\

S$2$SCORE$_T$-QE
& \textbf{0.659} & \textbf{0.653} &
  \textbf{0.644} & \textbf{0.629} &
  \textbf{0.650} & \textbf{0.647} \\
\midrule

\multicolumn{7}{l}{\textit{xCOMET-XXL (Training supervision)}} \\
\cmidrule(lr){1-7}
BLASER~$2.0$-REF
& 0.633 & 0.624 & 0.615 & 0.610 & 0.621 & 0.621 \\

BLASER~$2.0$-QE
& 0.657 & 0.649 & 0.620 & 0.619 & 0.642 & 0.637 \\

S$2$SCORE$_T$-REF
& 0.671 & 0.662 & 0.655 & 0.635 & 0.659 & 0.656 \\

S$2$SCORE$_T$-QE
& \textbf{0.686} & \textbf{0.673} &
  \textbf{0.663} & \textbf{0.641} &
  \textbf{0.670} & \textbf{0.667} \\
\bottomrule
\end{tabular}
\end{table}

\begin{figure}[!t]
\centering
\includegraphics[width=0.35\textwidth]{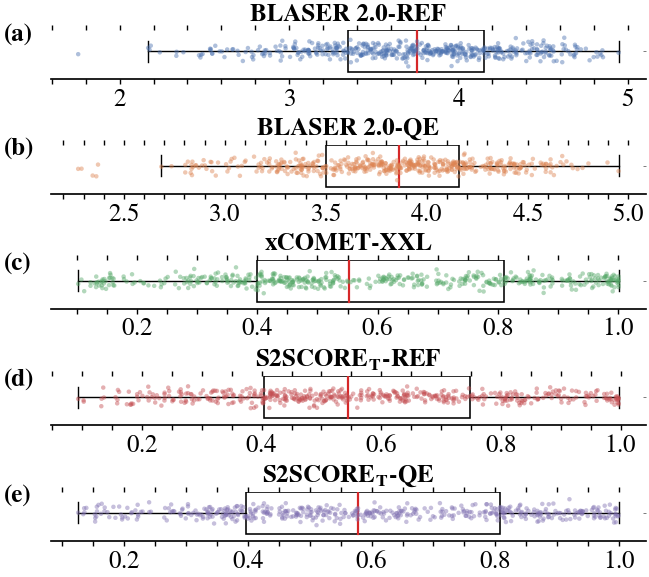}
\caption{Score distributions of semantic metrics for the fr--en test set. Panels: (a) BLASER~$2.0$-REF, (b) BLASER~$2.0$-QE, (c) xCOMET-XXL supervision, (d) S$2$SCORE$_T$-REF, and (e) S$2$SCORE$_T$-QE. Box plots show the median (red), interquartile range (box), and whiskers extending to $1.5\times\mathrm{IQR}$.}
\label{fig:score_dist}
\end{figure}

\subsection{Semantic Adequacy Evaluation}

We evaluate the semantic head using three complementary reference signals: human adequacy ratings, MetricX-$24$~\cite{juraska-etal-2024-metricx}, and xCOMET-XXL~\cite{xcomet}. Human ratings provide the primary assessment of semantic prediction quality, while MetricX-$24$ and xCOMET-XXL assess agreement with an independent semantic metric and the training supervision signal, respectively. As shown in Table~\ref{tab:semantic_eval}, S$2$SCORE$_T$ consistently achieves stronger agreement with human ratings than BLASER~$2.0$. S$2$SCORE$_T$-QE attains Spearman correlations of $0.607$--$0.649$ across the five language pairs, with a mean $\rho=0.631$, while S$2$SCORE$_T$-REF achieves correlations of $0.604$--$0.635$ (mean $\rho=0.621$). In comparison, BLASER~$2.0$-QE and BLASER~$2.0$-REF achieve mean correlations of $0.606$ and $0.597$, respectively. The improvements over the corresponding BLASER~$2.0$ baselines are statistically significant ($p<0.05$). Since human ratings are not used during training, these results indicate that the proposed speech-based metric better reflects human perceptions of semantic adequacy and generalizes across diverse language pairs, including zh--en.

S$2$SCORE$_T$ also shows strong agreement with MetricX-$24$, achieving correlations of $0.629$--$0.659$ (mean $\rho=0.647$), suggesting that the learned representation generalizes to an independent semantic metric. Agreement with xCOMET-XXL remains high, with correlations of $0.641$--$0.686$ (mean $\rho=0.667$), indicating consistency with the training objective.


Beyond rank correlation, Fig.~\ref{fig:score_dist} shows that S$2$SCORE$_T$ produces substantially wider score distributions than BLASER~$2.0$ on the fr--en test set. The normalized standard deviation increases from $0.119$--$0.144$ for BLASER~$2.0$ to $0.228$--$0.237$ for S$2$SCORE$_T$, approaching that of xCOMET-XXL ($0.251$). The broader distributions indicate greater discriminative capacity, enabling finer differentiation between translation outputs of varying semantic quality while operating directly on speech.

\begin{table}[!t]
\centering
\caption{Spearman correlation ($\rho$) of S$2$SCORE$_S$ predictions with human speaker-similarity ratings, ECAPA-TDNN~\cite{desplanques2020ecapa} cosine similarity~\cite{desplanques2020ecapa}, and WavLM-SV~\cite{chen2022wavlm} cosine similarity on the test set. Mean values are computed across five language pairs.}

\label{tab:speaker_all}
\setlength{\tabcolsep}{4pt}
\renewcommand{\arraystretch}{1.05}
\begin{tabular}{lcccccc}
\toprule
\textbf{Reference} &
\textbf{fr--en} &
\textbf{es--en} &
\textbf{de--en} &
\textbf{zh--en} &
\textbf{hi--en} &
\textbf{Mean} \\
\midrule
Human
& 0.857 & 0.864 & 0.859 & 0.862 & 0.870 & 0.862 \\

ECAPA-TDNN
& 0.861 & 0.873 & 0.868 & 0.867 & 0.893 & 0.872 \\

WavLM-SV
& 0.872 & 0.891 & 0.893 & 0.881 & 0.904 & 0.888 \\
\bottomrule
\end{tabular}
\end{table}

\begin{figure}[!t]
\centering
\includegraphics[width=0.38\textwidth]{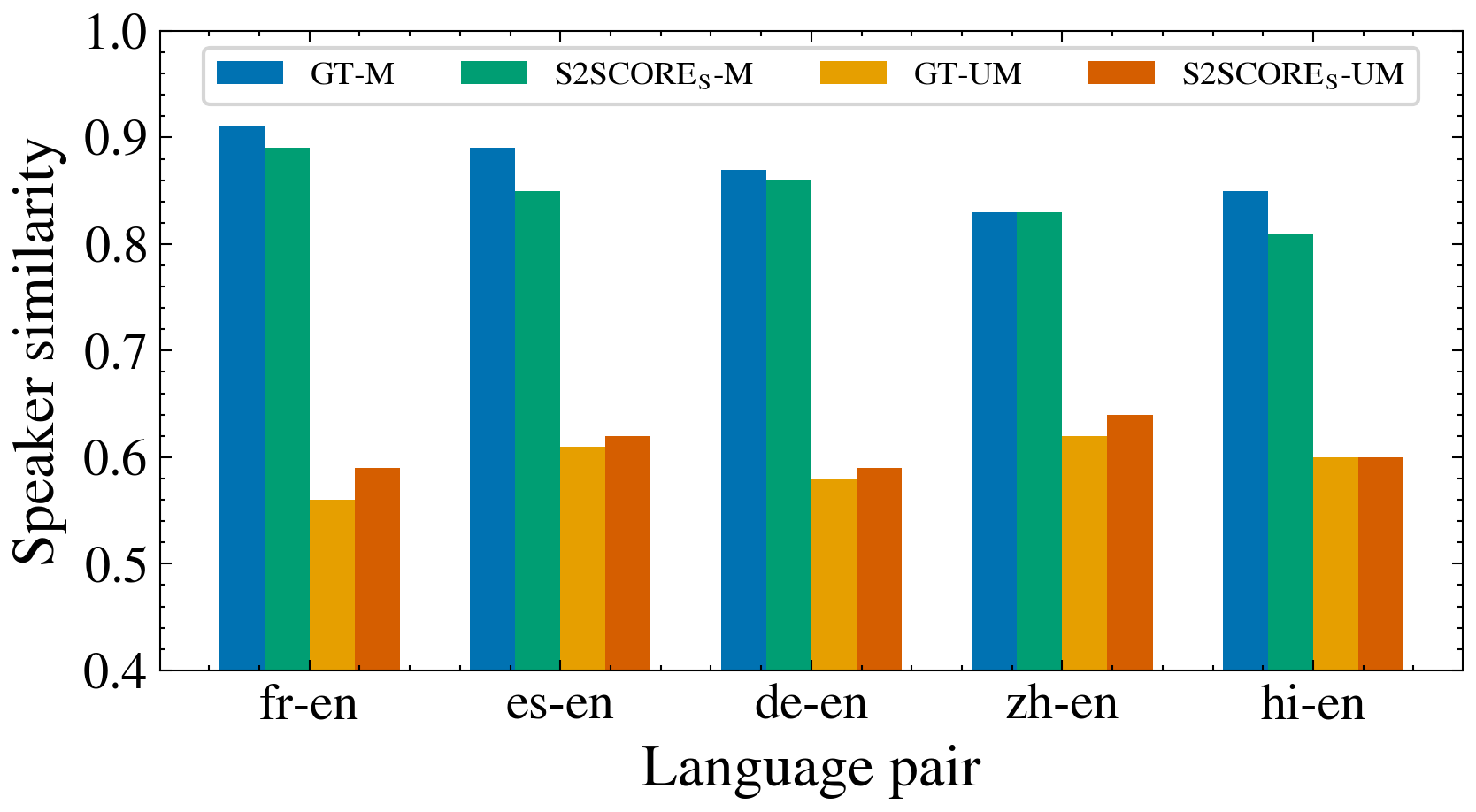}
\caption{Mean speaker similarity across language pairs. GT-M/GT-UM denote ground-truth WavLM-SV~\cite{chen2022wavlm} cosine similarities for matched/unmatched pairs; S$2$SCORE$_S$-M/S$2$SCORE$_S$-UM denote the corresponding predictions.}
\label{fig:spk_sim}
\end{figure}

\subsection{Speaker Similarity Evaluation}
\label{sec:speaker_eval}
We evaluate the speaker head against three complementary reference signals: human speaker-similarity ratings, ECAPA-TDNN~\cite{desplanques2020ecapa} cosine similarity, and WavLM-SV~\cite{chen2022wavlm} cosine similarity. Human ratings provide the primary assessment of speaker preservation, while ECAPA-TDNN and WavLM-SV measure agreement with an independent speaker verification model and the training supervision signal, respectively. As shown in Table~\ref{tab:speaker_all}, S$2$SCORE$_S$ achieves strong agreement with human speaker-similarity ratings across all language pairs, with Spearman correlations ranging from $0.857$ to $0.870$ (mean $\rho=0.862$). The consistently high correlations indicate that the learned representation effectively captures speaker preservation and generalizes across language pairs.

S$2$SCORE$_S$ also shows strong agreement with ECAPA-TDNN, achieving correlations of $0.861$--$0.893$ (mean $\rho=0.872$). Since ECAPA-TDNN is not used during training, these results indicate that the learned representation also aligns well with an independent speaker verification model. Agreement with the WavLM-SV supervision signal remains high, with correlations of $0.872$--$0.904$ (mean $\rho=0.888$), indicating strong consistency with the training objective.

Beyond rank correlation, Fig.~\ref{fig:spk_sim} shows that S$2$SCORE$_S$ preserves speaker-similarity discrimination across languages. Ground-truth WavLM-SV cosine similarities separate matched and unmatched speaker pairs, ranging from $0.831$ to $0.910$ and $0.562$ to $0.623$, respectively. S$2$SCORE$_S$ reproduces this pattern, with matched predictions of $0.810$--$0.890$ and unmatched predictions of $0.590$--$0.640$. The comparable separation margins in ground-truth similarities ($0.208$--$0.348$) and predictions ($0.170$--$0.300$) indicate that the metric retains the relative structure of speaker similarity across languages.


\begin{table}[!t]
\centering
\caption{Spearman correlation ($\rho$) of S$2$SCORE$_N$ predictions with human naturalness ratings, NISQA~\cite{mittag2021nisqa}, and UTMOSv$2$~\cite{baba2024utmosv2} scores on the test set. Mean values are computed across five language pairs.}
\label{tab:naturalness_all}
\setlength{\tabcolsep}{4pt}
\renewcommand{\arraystretch}{1.05}
\begin{tabular}{lcccccc}
\toprule
\textbf{Reference} &
\textbf{fr--en} &
\textbf{es--en} &
\textbf{de--en} &
\textbf{zh--en} &
\textbf{hi--en} &
\textbf{Mean} \\
\midrule
Human
& 0.890 & 0.901 & 0.883 & 0.851 & 0.871 & 0.879 \\

NISQA
& 0.899 & 0.918 & 0.892 & 0.875 & 0.882 & 0.893 \\

UTMOSv$2$
& 0.914 & 0.925 & 0.902 & 0.887 & 0.895 & 0.905 \\
\bottomrule
\end{tabular}
\end{table}

\begin{figure}[!t]
\centering
\includegraphics[width=0.40\textwidth]{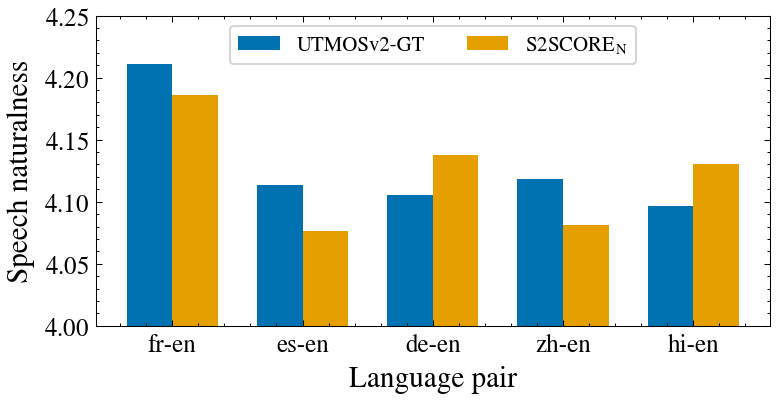}
\caption{Mean naturalness scores across language pairs. Blue bars denote UTMOSv$2$ reference scores and orange bars denote the corresponding S$2$SCORE$_N$ predictions on the $[1,5]$ MOS scale.}
\label{fig:nat_scores}
\end{figure}

\subsection{Acoustic Naturalness Evaluation}
\label{sec:naturalness_eval}

We evaluate the naturalness head against three complementary reference signals: human naturalness ratings, NISQA~\cite{mittag2021nisqa} predictions, and UTMOSv$2$~\cite{baba2024utmosv2} predictions. Human ratings provide the primary assessment of perceptual naturalness, while NISQA and UTMOSv$2$ measure agreement with an independent speech-quality model and the training supervision signal, respectively. As shown in Table~\ref{tab:naturalness_all}, S$2$SCORE$_N$ achieves strong agreement with human naturalness ratings across all language pairs, with Spearman correlations ranging from $0.851$ to $0.901$ (mean $\rho=0.879$). These results indicate that the model effectively captures perceptual naturalness and generalizes across languages. 

S$2$SCORE$_N$ also shows strong agreement with NISQA ($\rho=0.875$--$0.918$, mean $\rho=0.893$), suggesting generalization beyond a single speech-quality model. Agreement with UTMOSv$2$ remains similarly high ($\rho=0.887$--$0.925$, mean $\rho=0.905$), indicating characteristics learned from the supervision signal are largely preserved.

Beyond rank correlation, Fig.~\ref{fig:nat_scores} shows that S$2$SCORE$_N$ preserves the cross-lingual naturalness patterns of the underlying UTMOSv$2$ reference. Mean UTMOSv$2$ scores span $4.10$--$4.21$ on the $[1,5]$ MOS scale, while S$2$SCORE$_N$ predictions remain within $\pm0.04$ MOS of these values. S$2$SCORE$_N$ slightly underestimates naturalness for fr--en, es--en, and zh--en, and overestimates it for de--en and hi--en. Despite these minor offsets, S$2$SCORE$_N$ preserves both the dynamic range and cross-lingual variation of the naturalness signal.

\begin{table}[!t]
\centering
\caption{Spearman correlation ($\rho$) of unified S$2$SCORE predictions with human overall quality ratings on the test set. Mean values are computed across the five language pairs.}
\label{tab:human_unified}
\setlength{\tabcolsep}{4.5pt}
\renewcommand{\arraystretch}{1.05}
\begin{tabular}{lcccccc}
\toprule
\textbf{Model} &
\textbf{fr--en} &
\textbf{es--en} &
\textbf{de--en} &
\textbf{zh--en} &
\textbf{hi--en} &
\textbf{Mean} \\
\midrule
S$2$SCORE-REF
& 0.702 & 0.713 & 0.697 & 0.681 & 0.706 & 0.700 \\

S$2$SCORE-QE
& \textbf{0.718} & \textbf{0.726} &
  \textbf{0.711} & \textbf{0.694} &
  \textbf{0.719} & \textbf{0.714} \\
\bottomrule
\end{tabular}
\end{table}

\begin{table}[!t]
\centering
\caption{Unified S$2$SCORE values across language pairs under matched (M) and unmatched (UM) speaker conditions. Mean values are computed across the five language pairs.}
\label{tab:overall_scores}
\setlength{\tabcolsep}{5pt}
\renewcommand{\arraystretch}{1.05}
\begin{tabular}{lcccccc}
\toprule
\textbf{Model} &
\textbf{fr--en} &
\textbf{es--en} &
\textbf{de--en} &
\textbf{zh--en} &
\textbf{hi--en} &
\textbf{Mean} \\
\midrule
REF-M  & 0.787 & 0.760 & 0.767 & 0.749 & 0.753 & 0.763 \\
REF-UM & 0.687 & 0.684 & 0.677 & 0.685 & 0.683 & 0.683 \\
QE-M   & 0.792 & 0.764 & 0.771 & 0.752 & 0.758 & 0.767 \\
QE-UM  & 0.692 & 0.687 & 0.681 & 0.689 & 0.688 & 0.687 \\
\bottomrule
\end{tabular}
\end{table}

\subsection{Unified S$2$ST Quality Score}
\label{sec:unified_eval}
We evaluate unified S$2$SCORE using human overall quality ratings across language pairs and speaker conditions. As shown in Table~\ref{tab:human_unified}, S$2$SCORE-QE achieves Spearman correlations of $0.694$--$0.726$ (mean $\rho=0.714$), while S$2$SCORE-REF attains comparable performance ($0.681$--$0.713$, mean $\rho=0.700$). These results indicate that the unified score reflects human judgments across diverse translation and acoustic conditions. The QE variant performs slightly better, consistent with the semantic evaluation. This may be because source and translated speech capture most quality-relevant information, while the reference provides limited additional information relative to the increased modeling complexity.

Table~\ref{tab:overall_scores} further shows that matched-speaker pairs consistently receive higher unified scores than unmatched pairs, with an average margin of approximately $0.08$ for both REF and QE variants, highlighting speaker preservation. The learned weights assign the largest contribution to semantic adequacy ($\alpha_{\mathrm{T}}=0.47$), followed by speaker preservation ($\alpha_{\mathrm{S}}=0.28$) and acoustic naturalness ($\alpha_{\mathrm{N}}=0.25$).  These weights reflect the relative contribution of each dimension to overall human quality, with higher semantic weight indicating a greater contribution from translation correctness, while speaker preservation and naturalness provide complementary information.



\begin{table}[!t]
\centering
\caption{Effect of multi-task training. Spearman correlation ($\rho$) between predictions and human ratings on the test split, averaged across five language pairs.}
\label{tab:ablation_multitask}
\setlength{\tabcolsep}{6pt}
\renewcommand{\arraystretch}{1.05}
\begin{tabular}{lcccc}
\toprule
\textbf{Method} & \textbf{Semantic} & \textbf{Naturalness} & \textbf{Speaker} & \textbf{Mean} \\
\midrule
Single-task & 0.617 & 0.859 & 0.849 & 0.775 \\
Multi-task  & \textbf{0.631} & \textbf{0.879} & \textbf{0.862} & \textbf{0.791} \\
\bottomrule
\end{tabular}
\end{table}

\begin{table}[!t]
\centering
\caption{Aggregation function comparison. Spearman correlation ($\rho$) with human overall quality ratings on the test split, averaged across five language pairs.}
\label{tab:ablation_agg}
\setlength{\tabcolsep}{4pt}
\renewcommand{\arraystretch}{1.05}
\begin{tabular}{lccc}
\toprule
\textbf{Metric} &
\textbf{Unweighted Mean} &
\textbf{Learned Linear (Ours)} &
\textbf{MLP} \\
\midrule
Spearman $\rho$ & 0.671 & 0.714 & \textbf{0.720} \\
\bottomrule
\end{tabular}
\end{table}
\subsection{Ablation Studies}
\label{subsec:ablations}
To assess the effect of parameter sharing, Table~\ref{tab:ablation_multitask} compares the proposed multi-task framework with independently trained single-task models using Spearman correlation with human ratings on the test set. The multi-task framework consistently outperforms single-task models across all dimensions, with improvements of $+0.014$, $+0.020$, and $+0.013$ for semantic adequacy, acoustic naturalness, and speaker preservation, respectively, resulting in an average gain of $0.016$ Spearman correlation. Although modest, the consistent gains across all dimensions suggest that joint multi-task training improves the unified assessment.


To assess the aggregation function, Table~\ref{tab:ablation_agg} compares three strategies for the unified score: an equal-weight mean ($\alpha_{\mathrm{T}}=\alpha_{\mathrm{N}}=\alpha_{\mathrm{S}}=1/3$), a learned linear weighting in which the aggregation coefficients are optimized using human ratings, and a nonlinear two-layer MLP regressor. The learned weighting improves Spearman correlation by $0.043$ over the equal-weight mean, demonstrating the benefit of learning aggregation weights from human judgments. The two-layer MLP yields only a marginal gain of $0.006$ despite its additional parameters and complexity.

\section{Conclusion}
\label{sec:conclusions}
We introduced S$2$SCORE, a unified evaluation framework for S$2$ST that jointly models semantic adequacy, acoustic naturalness, and speaker similarity within a single architecture. Operating directly in the speech domain, S$2$SCORE enables holistic assessment of translated speech without requiring intermediate text representations. Across five language pairs, the proposed framework demonstrates strong agreement with human judgments and established reference metrics, while the unified score reliably predicts overall S$2$ST quality and generalizes across evaluated language pairs. The present study focuses on three key dimensions of S$2$ST quality and evaluates translation into English across five language pairs. Future work will extend the S$2$SCORE framework to incorporate additional aspects of spoken communication, such as prosody, emotional expressiveness, and speaking style, and evaluate its robustness across additional languages, translation directions, and more typologically diverse settings.

\bibliographystyle{IEEEtran}
\bibliography{mybib}

@inproceedings{papineni2002bleu,
  title     = {{BLEU}: a Method for Automatic Evaluation of Machine Translation},
  author    = {Papineni, Kishore and Roukos, Salim and Ward, Todd and Zhu, Wei-Jing},
  booktitle = {Proceedings of ACL},
  year      = {2002},
  pages     = {311--318},
}

@inproceedings{banerjee2005meteor,
  title     = {{METEOR}: An Automatic Metric for {MT} Evaluation with Improved Correlation with Human Judgments},
  author    = {Banerjee, Satanjeev and Lavie, Alon},
  booktitle = {Proceedings of the ACL Workshop},
  year      = {2005},
  pages     = {65--72},
}

@inproceedings{sellam2020bleurt,
  title     = {{BLEURT}: Learning Robust Metrics for Text Generation},
  author    = {Sellam, Thibault and Das, Dipanjan and Parikh, Ankur},
  booktitle = {Proceedings of ACL},
  year      = {2020},
  pages     = {7881--7892},
}

@inproceedings{rei2020comet,
  title     = {{COMET}: A Neural Framework for {MT} Evaluation},
  author    = {Rei, Ricardo and Stewart, Craig and Farinha, Ana C and Lavie, Alon},
  booktitle = {Proceedings of EMNLP},
  year      = {2020},
  pages     = {2685--2702},
}

@inproceedings{blaser,
  title     = {{BLASER}: A Text-Free Speech-to-Speech Translation Evaluation Metric},
  author    = {Chen, Mingda and Duquenne, Paul-Ambroise and Andrews, Pierre and Kao, Justine and Mourachko, Alexandre and Schwenk, Holger and Costa-juss{\`a}, Marta R.},
  booktitle = {Proceedings of ACL},
  year      = {2023},
  pages     = {9064--9079},
}

@inproceedings{blaser2,
  title     = {{BLASER} 2.0: A Metric for Evaluation and Quality Estimation of Massively Multilingual Speech and Text Translation},
  author    = {Dale, David and Costa-juss{\`a}, Marta R.},
  booktitle = {Findings of EMNLP},
  year      = {2024},
  pages     = {16075--16085},
}

@misc{besacier2023textless,
  author = {Laurent Besacier and Swen Ribeiro and Olivier Galibert and Ioan Calapodescu},
  title  = {A Textless Metric for Speech-to-Speech Comparison},
  year   = {2022},
  note   = {arXiv:2210.11835},
}

@inproceedings{speechqe,
  title     = {{SpeechQE}: Estimating the Quality of Direct Speech Translation},
  author    = {Han, HyoJung and Duh, Kevin and Carpuat, Marine},
  booktitle = {Proceedings of EMNLP},
  year      = {2024},
  pages     = {21852--21867},
}

@article{dhawan2022s2st,
  author  = {Dhawan, Sandeep},
  title   = {Speech to Speech Translation: Challenges and Future},
  journal = {IJCATR},
  volume  = {11},
  number  = {3},
  pages   = {36--55},
  year    = {2022},
}

@Article{Etchego2022,
AUTHOR = {Etchegoyhen, Thierry and Arzelus, Haritz and Gete, Harritxu and Alvarez, Aitor and Torre, Iván G. and Martín-Doñas, Juan Manuel and González-Docasal, Ander and Fernandez, Edson Benites},
TITLE = {Cascade or Direct Speech Translation? A Case Study},
JOURNAL = {Applied Sciences},
VOLUME = {12},
YEAR = {2022},
NUMBER = {3},
ARTICLE-NUMBER = {1097}
}

@ARTICLE{alharbi2021asr,
  author={Alharbi, Sadeen and Alrazgan, Muna and Alrashed, Alanoud and Alnomasi, Turkiayh and Almojel, Raghad and Alharbi, Rimah and Alharbi, Saja and Alturki, Sahar and Alshehri, Fatimah and Almojil, Maha},
  journal={IEEE Access}, 
  title={Automatic Speech Recognition: Systematic Literature Review}, 
  year={2021},
  volume={9},
  pages={131858-131876}}

@article{tan2020nmt,
  author  = {Zhixing Tan and Shuo Wang and Zonghan Yang and Gang Chen and Xuancheng Huang and Maosong Sun and Yang Liu},
  title   = {Neural Machine Translation: A Review of Methods, Resources, and Tools},
  journal = {AI Open},
  volume  = {1},
  pages   = {5--21},
  year    = {2020}
}

@inproceedings{casanova2024xtts,
  author    = {Edresson Casanova and Kelly Davis and Eren G{\"o}lge and G{\"o}rkem G{\"o}knar and Iulian Gulea and Logan Hart and Aya Aljafari and Joshua Meyer and Reuben Morais and Samuel Olayemi and Julian Weber},
  title     = {{XTTS}: a Massively Multilingual Zero-Shot Text-to-Speech Model},
  booktitle = {Proceedings of Interspeech},
  pages     = {4978--4982},
  year      = {2024},
}

@inproceedings{le2024transvip,
title={Trans{VIP}: Speech to Speech Translation System with Voice and Isochrony Preservation},
author={Chenyang Le and Yao Qian and Dongmei Wang and Long Zhou and Shujie LIU and Xiaofei Wang and Midia Yousefi and Yanmin Qian and Jinyu Li and Michael Zeng},
booktitle = {Proceedings of NeurIPS},
year={2024},
pages={89682--89705}
}

@inproceedings{jia2019direct,
  author    = {Ye Jia and Ron J. Weiss and Fadi Biadsy and Wolfgang Macherey and Melvin Johnson and Zhifeng Chen and Yonghui Wu},
  title     = {Direct Speech-to-Speech Translation with a Sequence-to-Sequence Model},
  booktitle = {Proceedings of Interspeech},
  year      = {2019},
  pages     = {1123--1127},
}

@article{xcomet,
  title   = {x{COMET}: Transparent Machine Translation Evaluation through Fine-grained Error Detection},
  author  = {Guerreiro, Nuno M. and Rei, Ricardo and van Stigt, Daan and Coheur, Luisa and Colombo, Pierre and Martins, Andr{\'e} F. T.},
  journal = {TACL},
  volume  = {12},
  pages   = {979--995},
  year    = {2024},
}

@misc{duquenne2023sonar,
  author = {Paul-Ambroise Duquenne and Holger Schwenk and Beno{\^\i}t Sagot},
  title  = {{SONAR}: Sentence-Level Multimodal and Language-Agnostic Representations},
  year   = {2023},
  eprint = {2308.11466},
  note   = {arXiv:2308.11466},
}

@article{chen2022wavlm,
  author  = {Chen, Sanyuan and Wang, Chengyi and Chen, Zhengyang and Wu, Yu and Liu, Shujie and Chen, Zhuo and Li, Jinyu and Kanda, Naoyuki and Yoshioka, Takuya and Xiao, Xiong and Wu, Jian and Zhou, Long and Ren, Shuo and Qian, Yanmin and Qian, Yao and Wu, Jian and Zeng, Michael and Yu, Xiangzhan and Wei, Furu},
  title   = {Wav{LM}: Large-Scale Self-Supervised Pre-Training for Full Stack Speech Processing},
  journal = {IEEE J. Sel. Topics Signal Process.},
  volume  = {16},
  number  = {6},
  pages   = {1505--1518},
  year    = {2022},
}

@article{davis1980comparison,
  title={Comparison of parametric representations for monosyllabic word recognition in continuously spoken sentences},
  author={Davis, Steven B. and Mermelstein, Paul},
  journal={IEEE Transactions on Acoustics, Speech, and Signal Processing},
  volume={28},
  number={4},
  pages={357--366},
  year={1980}
}

@misc{oord2018representation,
  author = {Aaron van den Oord and Yazhe Li and Oriol Vinyals},
  title  = {Representation Learning with Contrastive Predictive Coding},
  year   = {2019},
  eprint = {1807.03748},
  note   = {arXiv:1807.03748},
}

@inproceedings{jia2022cvss,
  title     = {{CVSS} Corpus and Massively Multilingual Speech-to-Speech Translation},
  author    = {Jia, Ye and Tadmor Ramanovich, Michelle and Wang, Quan and Zen, Heiga},
  booktitle = {Proceedings of LREC},
  year      = {2022},
  pages     = {6691--6703},
}

@inproceedings{jain2025bhasaanuvaad,
  author    = {Ashwin Sankar and Sparsh Jain and Nikhil Narasimhan and Devilal Choudhary and Dhairya Suman and Mohammed Safi Ur Rahman Khan and Anoop Kunchukuttan and Mitesh M. Khapra and Raj Dabre},
  title     = {Towards Building Large Scale Datasets and State-of-the-Art Automatic Speech Translation Systems for 14 Indian Languages},
  booktitle = {Proceedings of ACL},
  pages     = {32945--32966},
  year      = {2025},
}

@inproceedings{radford2023whisper,
  author    = {Radford, Alec and Kim, Jong Wook and Xu, Tao and Brockman, Greg and McLeavey, Christine and Sutskever, Ilya},
  title     = {Robust Speech Recognition via Large-Scale Weak Supervision},
  booktitle = {Proceedings of ICML},
  year      = {2023},
  pages     = {28492--28518},
}

@misc{barrault2023seamlessm4t,
  author = {Seamless Communication and Loïc Barrault and Yu-An Chung and Mariano Cora Meglioli and David Dale and Ning Dong and Paul-Ambroise Duquenne and Hady Elsahar and Hongyu Gong and Kevin Heffernan and John Hoffman and Christopher Klaiber and Pengwei Li and Daniel Licht and Jean Maillard and Alice Rakotoarison and Kaushik Ram Sadagopan and Guillaume Wenzek and Ethan Ye and Bapi Akula and Peng-Jen Chen and Naji El Hachem and Brian Ellis and Gabriel Mejia Gonzalez and Justin Haaheim and Prangthip Hansanti and Russ Howes and Bernie Huang and Min-Jae Hwang and Hirofumi Inaguma and Somya Jain and Elahe Kalbassi and Amanda Kallet and Ilia Kulikov and Janice Lam and Daniel Li and Xutai Ma and Ruslan Mavlyutov and Benjamin Peloquin and Mohamed Ramadan and Abinesh Ramakrishnan and Anna Sun and Kevin Tran and Tuan Tran and Igor Tufanov and Vish Vogeti and Carleigh Wood and Yilin Yang and Bokai Yu and Pierre Andrews and Can Balioglu and Marta R. Costa-jussà and Onur Celebi and Maha Elbayad and Cynthia Gao and Francisco Guzmán and Justine Kao and Ann Lee and Alexandre Mourachko and Juan Pino and Sravya Popuri and Christophe Ropers and Safiyyah Saleem and Holger Schwenk and Paden Tomasello and Changhan Wang and Jeff Wang and Skyler Wang},
  title  = {{SeamlessM4T}: Massively Multilingual \& Multimodal Machine Translation},
  year   = {2023},
  eprint = {2308.11596},
  note   = {arXiv:2308.11596},
}

@misc{arya2026timbreaware,
  author = {Lalaram Arya and Mrinmoy Bhattacharjee and Adarsh C. R. and S. R. Mahadeva Prasanna},
  title  = {Timbre-Aware {LLM}-based Direct Speech-to-Speech Translation Extendable to Multiple Language Pairs},
  year   = {2026},
  eprint = {2601.16023},
  note   = {arXiv:2601.16023},
}

@article{Yu2025Multitask,
  author  = {Yu, Jun and Liu, Xiaokang and Luo, Chongliang and Zhou, Rong and Liu, Yixin and Hu, Jie and Chen, Jianmin and Zhang, Ke and Zhang, Dazheng and Shen, Yishan and Adhikarla, Eashan and Dai, Yutong and Zhang, Kai and Kong, Zhaoming and Ye, Wenxuan and Yin, Yilong and Namboodiri, Vinod and Davison, Brian and Moore, Jason and Chen, Yong},
  title   = {Multitask Learning 1997--2024: Part I Fundamentals},
  journal = {Harvard Data Science Review},
  volume  = {7},
  number  = {3},
  year    = {2025},
}

@inproceedings{loshchilov2019adamw,
  author    = {Ilya Loshchilov and Frank Hutter},
  title     = {Decoupled Weight Decay Regularization},
  booktitle = {Proceedings of ICLR},
  year      = {2019},
}

@inproceedings{juraska-etal-2024-metricx,
  title     = "{M}etric{X}-24: The {G}oogle Submission to the {WMT} 2024 Metrics Shared Task",
  author    = {Juraska, Juraj and Deutsch, Daniel and Finkelstein, Mara and Freitag, Markus},
  booktitle = {Proceedings of WMT},
  year      = {2024},
  pages     = {492--504},
}

@inproceedings{baba2024utmosv2,
  title     = {The {T}05 System for The {V}oice{MOS} {C}hallenge 2024: Transfer Learning from Deep Image Classifier to Naturalness {MOS} Prediction of High-Quality Synthetic Speech},
  author    = {Baba, Kaito and Nakata, Wataru and Saito, Yuki and Saruwatari, Hiroshi},
  booktitle = {Proceedings of SLT},
  year      = {2024},
}

@inproceedings{desplanques2020ecapa,
  author    = {Brecht Desplanques and Jenthe Thienpondt and Kris Demuynck},
  title     = {{ECAPA-TDNN}: Emphasized Channel Attention, Propagation and Aggregation in {TDNN} Based Speaker Verification},
  booktitle = {Proceedings of Interspeech},
  pages     = {3830--3834},
  year      = {2020},
}

@inproceedings{mittag2021nisqa,
  author    = {Mittag, Gabriel and Naderi, Babak and Chehadi, A. and Möller, Sebastian},
  title     = {{NISQA}: A Deep CNN-Self-Attention Model for Multidimensional Speech Quality Prediction with Crowdsourced Datasets},
  booktitle = {Proceedings of Interspeech},
  pages     = {2127--2131},
  year      = {2021},
}

@inproceedings{juraska2023metricx,
  author    = {Juraska, Juraj and Finkelstein, Mara and Deutsch, Daniel and Siddhant, Aditya and Mirzazadeh, Mehdi and Freitag, Markus},
  title     = {{M}etric{X}-23: The {G}oogle Submission to the {WMT} 2023 Metrics Shared Task},
  booktitle = {Proceedings of WMT},
  pages     = {756--767},
  year      = {2023},
}

@inproceedings{koehn2004statistical,
  author    = {Koehn, Philipp},
  title     = {Statistical Significance Tests for Machine Translation Evaluation},
  booktitle = {Proceedings of EMNLP},
  pages     = {388--395},
  year      = {2004},
}

@article{spearman1904proof,
  author  = {Spearman, Charles},
  title   = {The Proof and Measurement of Association between Two Things},
  journal = {The American Journal of Psychology},
  volume  = {15},
  number  = {1},
  pages   = {72--101},
  year    = {1904},
}

\end{document}